\documentstyle[11pt,newpasp,twoside,epsf]{article}

\def\edcomment#1{\iffalse\marginpar{\raggedright\sl#1\/}\else\relax\fi}
\reversemarginpar

\def\MLF{{\sf MLF}}
\def\lax {\>\vcenter{\hbox{$<$\hskip-.75em\lower1.0ex\hbox{$\sim$}}}\>}
\def\gax {\>\vcenter{\hbox{$>$\hskip-.75em\lower1.0ex\hbox{$\sim$}}}\>}
\def\ion#1#2{#1~{\sc\romannumeral #2}}

\begin{document}

\title{The Fe XXII I(11.92~\AA )/I(11.77~\AA ) Density Diagnostic}
\author{Christopher W.\ Mauche, Duane A.\ Liedahl, and Kevin B.\ Fournier}
\affil{Lawrence Livermore National Laboratory,\\
L-473, 7000 East Avenue, Livermore, CA 94550}

\begin{abstract}
Using the Livermore X-ray Spectral Synthesizer, which calculates spectral
models of highly charged ions based on HULLAC atomic data, we investigate
the temperature, density, and photoexcitation dependence of the $I(11.92~{\rm
\AA })/I(11.77~{\rm\AA })$ line ratio of \ion{Fe}{22}. Applied to the {\it
Chandra\/} HETG spectrum of the intermediate polar EX~Hya, we find that the
electron density of its $T_{\rm e}\approx 12$ MK plasma is $\log n_{\rm e}\,
({\rm cm^{-3}}) =14.3^{+0.7}_{-0.5}$, orders of magnitude greater than that
observed in the Sun or other late-type stars.
\end{abstract}

\section{Introduction}

The unique aspect of the high-temperature plasma in the post-shock region of
the accretion flow of magnetic cataclysmic variables (mCVs) is its density,
which is expected to be high because of the magnetic funneling of the mass
lost by the secondary; the factor-of-four density jump across the accretion
shock; and the settling nature of the post-shock flow, where the density
scales inversely with the temperature. The standard density diagnostic of
high-temperature plasmas is the intensity ratio $R\equiv f/i$ of the forbidden
to intercombination lines of He-like ions, but this diagnostic is compromised
in mCVs for two reasons (see Mauche 2002 for details). First, the critical
density of this ratio increases with $Z$ and hence temperature---the opposite
of the trend in the accretion column---so the $R$ line ratio is effective over
only a narrow range of temperatures. Second, in mCVs and other UV-bright
stars, photoexcitation competes with collisional excitation to depopulate the
upper level of the forbidden line, so the $R$ line ratio can appear to be in
the ``high-density limit'' if the radiation field is sufficiently strong at
the appropriate wavelengths. For example, in a plasma illuminated by a 30 kK
blackbody, the $R$ line ratios of all elements through Mg lie in the
high-density limit, regardless of the density.

Recently, Mauche, Liedahl, \& Fournier (2001, hereafter \MLF ) discussed the
temperature, density, and photoexcitation sensitivity of the $I(17.10~{\rm
\AA })/$ $I(17.05~{\rm\AA })$ line ratio of \ion{Fe}{17}. This line ratio is
well suited to mCVs because its critical density is high ($n_{\rm c}\approx
3\times 10^{13}~\rm cm^{-3}$, comparable to that of \ion{Si}{13}) and it is
less sensitive to photoexcitation. Applied to the {\it Chandra\/} HETG
spectrum of the intermediate polar EX~Hya, \MLF \ showed that the measured
$I(17.10~{\rm\AA })/I(17.05~{\rm \AA })$ line ratio can be explained if the
electron density $n_{\rm e} \gax 2\times 10^{14}~\rm cm^{-3}$, or if the
photoexcitation temperature $T_{\rm bb}\gax 55$~kK.

\begin{figure}
\plotone{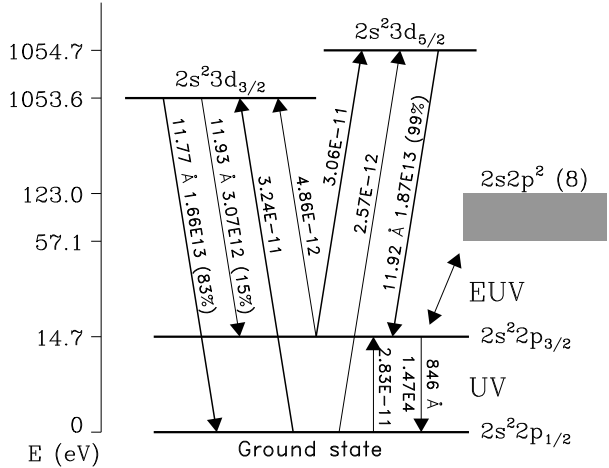} 
\caption{Schematic energy level diagram of \ion{Fe}{22} showing the 
wavelengths,
radiative decay rates, and branching ratios ({\it downward pointing arrows\/})
and collisional rate coefficients for $T_{\rm e}=12.8$ MK ({\it upward pointing
arrows\/}) for the $2p$--$3d$ transitions. Population in the $2s^22p_{3/2}$
level is also built up through radiative cascades from the $2s2p^2$ manifold.}
\vskip -4pt
\end{figure}

\section{Fe XXII}

Seeking to circumvent the density/photoexcitation ambiguity of the He-like $R$
and \ion{Fe}{17} $I(17.10~{\rm\AA })/I(17.05~{\rm \AA })$ line ratios, we
undertook an investigation of the temperature, density, and photoexcitation
dependence of the $3\rightarrow 2$ transitions of \ion{Fe}{22} (see Wargelin et
al.\ 1998 for a discussion of the \ion{Fe}{22} $4\rightarrow 2$ transitions).
Figure~1 shows a schematic of the important level-population processes in
\ion{Fe}{22}. The density sensitivity of the $n\rightarrow 2$ transitions of
\ion{Fe}{22} is due to the build-up of electron population in the first excited
$2s^22p_{3/2}$ level due to the slow radiative decay rate out of that level,
the 15\% branching ratio from the $2s^23d_{3/2}$ level, and cascades 
through the
$2s2p^2$ manifold. The population in the $2s^22p_{3/2}$ level is essentially
unaffected by photoexcitation because the $2s^22p_{1/2}$--$2s^22p_{3/2}$
transition in the UV is not optically allowed, and the $2s^22p_{3/2}$--$2s2p^2$
transitions lie in the EUV, where observations (Hurwitz et al.\ 1997)
demonstrate there is at most a weak source of continuum photons. The density
sensitivity of the $3 \rightarrow 2$ transitions of \ion{Fe}{22} manifests
itself most clearly in the $I(11.92~{\rm \AA })/I(11.77~{\rm\AA })$ line ratio.
This ratio is $\sim 0.4$ in low-density plasma (Brown et al.\ 2002), but is
a factor of 2--3 times higher---$1.06\pm 0.23$---in the {\it Chandra\/} HETG
spectrum of EX~Hya (see Fig.~2).

\begin{figure}
\plotone{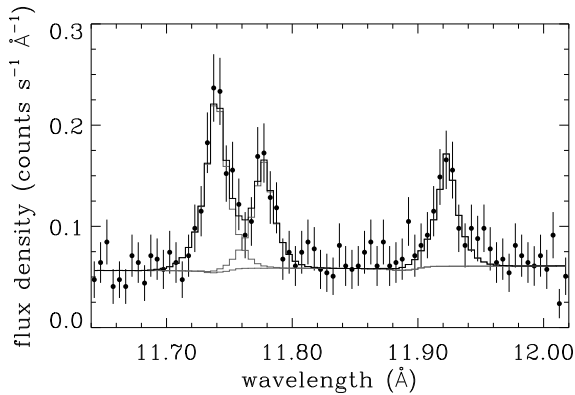} 
\caption{Detail of the {\it Chandra\/} MEG spectrum of EX~Hya in the
neighborhood of the \ion{Fe}{22} $2p$--$3d$ lines. Data are shown by the filled
circles with error bars and the model (3 Gaussians plus a linear background
convolved with the HETG spectral resolution and effective area) is shown by
the histograms. The line at 11.74~\AA \ is due to \ion{Fe}{23}.}
\vskip -4pt
\end{figure}

To investigate the atomic kinetics of the $3\rightarrow 2$ transitions of
\ion{Fe}{22}, we used the Livermore X-ray Spectral Synthesizer (LXSS), a suite
of IDL codes that calculates spectral models of highly charged ions based on
Hebrew University/Lawrence Livermore Atomic Code (HULLAC) atomic data. Our
\ion{Fe}{22} model includes radiative transition rates for E1, E2, M1, and
M2 decays and electron impact excitation rate coefficients for levels with
principal quantum number $n\le 5$ and azimuthal quantum number $l\le 4$ for a
total of 228 levels. To account for photoexcitation, we included in the LXSS
population kinetics calculation the photoexcitation rates $(\pi e^2/m_ec)
f_{ij} (4\pi /h\nu ) B_\nu (T_{\rm bb})$, where $B_\nu (T_{\rm bb})$ is the
blackbody spectral energy distribution and $f_{ij}$ are the oscillator
strengths of the various transitions. Using these data, LXSS calculates the
level populations for a given temperature and density assuming
collisional-radiative equilibrium; the line intensities are then simply the
product of the level populations and the radiative decay rates (for details,
see \MLF ).

\begin{figure}
\plotone{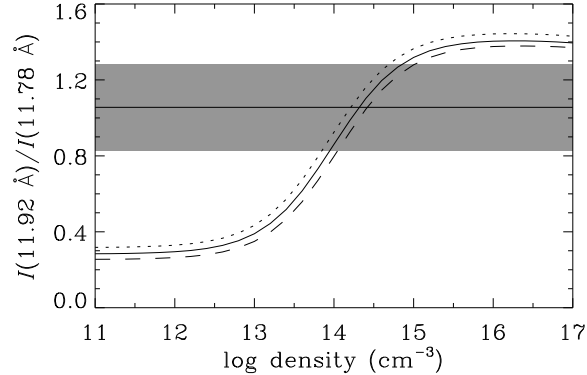} 
\caption{LXSS model of the \ion{Fe}{22} $I(11.92~{\rm \AA })/I(11.77~{\rm\AA})$
line ratio as a function of electron density for temperatures $T_{\rm e}=6.3$,
12.8, and 25.5 MK ({\it dotted, solid, and dashed curves, respectively\/}). The
horizontal line and shaded stripe indicate the value and $1\, \sigma$ error
envelope of the line ratio measured in EX~Hya.}
\vskip -4pt
\end{figure}

Figure~3 shows the results of our calculations for the \ion{Fe}{22}
$I(11.92~{\rm \AA })/$ $I(11.77~{\rm\AA })$ line ratio as a function of
electron density for temperatures $T_{\rm e}=6.3$, 12.8, and 25.5 MK. We find
that this line ratio has a critical density $n_{\rm c}\approx 10^{14}~\rm
cm^{-3}$, is $\approx 0.3$ at low densities and $\approx 1.4$ at high
densities, has only a very weak temperature dependence, and has only a weak
photoexcitation dependence for blackbodies with temperatures $T_{\rm bb}\lax
60$ kK. Given the measured value of the $I(11.92~{\rm \AA })/I(11.77~{\rm \AA
})$ line ratio in EX~Hya, we infer that the electron density of its $T_{\rm e}
\approx 12$ MK plasma is $\log n_{\rm e}~({\rm cm^{-3}}) =14.3^{+0.7}_{-0.5}$
(see Fig.~3).

\section{Summary and Conclusions}

To date, four spectroscopic diagnostics have provided evidence of 
high densities
in the X-ray--emitting plasma of the intermediate polar EX~Hya. Hurwitz et al.\
(1997) used the line ratio of the \ion{Fe}{20}/\ion{Fe}{23} 133~\AA \ blend to
the \ion{Fe}{21} 129~\AA \ line observed in the 1994 {\it EUVE\/} spectrum of
EX~Hya to infer $n_{\rm e}\gax 10^{13}~\rm cm^{-3}$ for its $T_{\rm e}\approx
10$~MK plasma. Using the 2000 {\it Chandra\/} HETG spectrum of EX~Hya, Mauche
(2002) showed that the He-like $R$ line ratios of O, Ne, Mg, Si, and S are
all in their high-density limit; \MLF \ used the \ion{Fe}{17}
$I(17.10~{\rm\AA })/I(17.05~{\rm \AA })$ line ratio to infer $n_{\rm e}\gax
2\times 10^{14}~\rm cm^{-3}$ for its $T_{\rm e}\approx 4$~MK plasma; and we
here used the \ion{Fe}{22} $I(11.92~{\rm\AA })/I(11.77~{\rm\AA })$ line ratio
to infer $n_{\rm e}\approx 2\times 10^{14}~\rm cm^{-3}$ for its $T_{\rm e}
\approx 12$ MK plasma. Of these diagnostics, the \ion{Fe}{17} and \ion{Fe}{22}
line ratios are the most reliable, with the \ion{Fe}{22} diagnostic limited
only by the quality of the data. We conclude first that the {\it Chandra\/}
HETG spectrum of EX~Hya requires plasma densities that are orders of magnitude
greater than those observed in the Sun or other late-type stars, and 
second that
the \ion{Fe}{17} $I(17.10~{\rm\AA })/I(17.05~{\rm\AA })$ and (better still)
the \ion{Fe}{22} $I(11.92~{\rm\AA })/I(11.77~{\rm\AA })$ line ratios provide
density diagnostics that are useful for sources like mCVs in which the plasma
densities are high and the efficacy of the He-like density diagnostics is
compromised by the presence of a bright UV continuum.

\acknowledgments

We thank H.~Tananbaum for the generous grant of Director's Discretionary Time
that made possible the {\it Chandra\/} observations of EX~Hya. Support for
this work was provided in part by NASA through {\it Chandra\/} Award Number
DD0-1004B issued by the {\it Chandra\/} X-Ray Observatory Center, which is
operated by the Smithsonian Astrophysical Observatory for and on behalf of NASA
under contract NAS8-39073. This work was performed under the auspices of the
U.S.~Department of Energy by University of California Lawrence Livermore
National Laboratory under contract No.~W-7405-Eng-48.


\end{document}